\documentclass[preprints,article,accept,moreauthors,pdftex]{Definitions/mdpi} 
\firstpage{1} 
\pubvolume{xx}
\issuenum{1}
\articlenumber{5}
\pubyear{2020}
\copyrightyear{2020}
\history{ }

\Title{Two-excitation routing via linear quantum channels}

\Author{Tony John George Apollaro $^{1,\dagger,\ddagger}$*\orcidA{0000-0002-9324-9336} and Wayne Jordan Chetcuti $^{2,3,4,\ddagger}$\orcidB{0000-0003-4676-8827} }

\AuthorNames{Tony John George Apollaro, Wayne Jordan Chetcuti}

\address{%
$^{1}$ \quad Department of Physics, Faculty of Science, University of Malta, Msida MSD 2080, Malta; tony.apollaro@um.edu.mt\\
$^{2}$ \quad Dipartimento di Fisica e Astronomia, Via S. Sofia 64, 95127 Catania, Italy; wayne.chetcuti@dfa.unict.it \\
$^{3}$ \quad INFN-Sezione di Catania, Via S. Sofia 64, 95127 Catania, Italy\\
$^{4}$ \quad Quantum Research Centre, Technology Innovation Institute, Abu Dhabi, UAE}

\corres{Correspondence: tony.apollaro@um.edu.mt; Tel.: +356-2340-8101}

\secondnote{These authors contributed equally to this work.}

\abstract{Routing quantum information among different nodes in a network is a fundamental prerequisite for a quantum internet. While single-qubit routing has been largely addressed, many-qubit routing protocols have not been intensively investigated so far. Building on the many-excitation transfer protocol in Ref.~\cite{Chetcuti2020}, we apply the perturbative transfer scheme to a two-excitation routing protocol on a network where multiple two-receivers block are coupled to a linear chain. We address both the case of switchable and permanent couplings between the receivers and the chain.
We find that the protocol allows for efficient two-excitation routing on a fermionic network, although for a spin-$\frac{1}{2}$ network only a limited region of the network is suitable for high-quality routing.}


\newcommand{\ket}[1]{\left\vert#1\right\rangle}
\newcommand{\bra}[1]{\left\langle#1\right\vert}

\newcommand{\opkb}[2]{\left\vert#1\right\rangle\!\!\left\langle#2\right\vert}

\newcommand{\average}[1]{\left\langle#1\right\rangle}

\begin{document}

\section{Introduction}
The coherent transfer of excitations from a sender to a receiver, located at different positions in a network, is of primary importance for many quantum-based technological applications, ranging from spintronics and atomtronics~\cite{Amico2020} to quantum-information processing~\cite{Nikolopoulos2014}. 

While a great amount of work has been devoted to the routing of the quantum state of a single qubit~\cite{Paganelli2013,Yousefjani2020a,Bayat2010,Yousefjani2020,Zueco2009,Chen2020,Wanisch2020, Pemberton-Ross2011,Zhan2014}, where the fidelity of the transfer protocol can be expressed in terms of the transition amplitude of a single excitation between a sender and a receiver location~\cite{PhysRevLett.91.207901}, the routing of a multiple qubit state is a far less investigated scenario. Although several protocols have been proposed both for two-qubit and multi-partite entangled quantum state transfer~\cite{Apollaro2015,doi:10.1142/S021974991750037X,qst2,Sousa2014,Vieira2019,VIEIRA20182586,Vieira2020,Apollaro2020,Almeida2019,Verma2017a}, their extension to a routing configuration on an arbitrary network is not straightforward. One reason being that almost all the proposed protocols rely on the quantum channel possessing mirror-symmetry, which, allowing for multiple receivers at arbitrary positions, is difficult to attain: in Ref.~\cite{Kay2011} it has been shown, e.g., that perfect state routing between multiple sites with real Hamiltonians is impossible. Moreover, the presence of a sender and a receiver block located at  positions other than the edges of a 1D quantum channel, implies that the total system is  no longer one-dimensional and the fermionisation of the spin chain via the celebrated Jordan-Wigner mapping is not valid anymore~\cite{Lieb1961}. As a consequence, the full spectrum of the network's Hamiltonian has to be found in the Hilbert space sector with two excitations and this can become, for long chains, quite cumbersome.

In this work we investigate the routing of two excitations by means of a linear chain, acting as a quantum wire, to which receivers can connect at arbitrary positions. Following the results of our recent work~\cite{Chetcuti2020}, we apply the weak-coupling protocol in order to route fermionic excitations on a 2D network. We consider both the case of switchable and permanent couplings of the receiver block to the quantum wire, obtaining the receivers' locations which allow for perturbatively perfect two-excitation transfer. We then compare the routing performance of fermions, which due to the non-interacting nature of the Hamiltonian considered in our work, can be analysed in terms of single-particle transition amplitudes, to the case where the network hosts spin-$\frac{1}{2}$ particles interacting via the $XX$-Heisenberg type Hamiltonian. We find that, although, a rigorous mapping of spins to non-interacting fermions is not possible because of the 2D nature of the network, it is indicated that several features of the free fermions dynamics can be retrieved also in the spin dynamics.

The paper is organised as follows. In Sec.~\ref{Sec.MBD} a brief introduction to the many-body dynamics in non-interacting fermion systems on a discrete lattice is given; in Sec.~\ref{Sec.1} the proposed protocol of two-excitation routing, both with switchable and permanent couplings, on a 2D lattice is presented; in Sec.~\ref{S.Spin} we analyse the case of spin-$\frac{1}{2}$ particle occupying the lattice positions of the network. Finally in Sec.~\ref{S.Dis} we discuss the main findings of our research and outline some future directions.

\section{Many-body dynamics in non-interacting fermions on a discrete lattice}\label{Sec.MBD}

Let us consider a discrete lattice model where each site can host one spinless fermion and whose dynamics is governed by the hopping Hamiltonian
\begin{equation}
\label{E.Ham_fer}
\hat{H}=\sum_{\average{ij}} J_{ij}\left(\hat{c}^{\dagger}_i\hat{c}_j+\textit{h.c}\right)~,
\end{equation}
where $\hat{c}^{\dagger}_i$ ($\hat{c}_i$) is the creation (annihilation) operator of a fermion on site $i$ and $J_{ij}$ is the kinetic term accounting for the hopping of a fermion between neighboring sites $i$ and $j$. This Hamiltonian conserves the total number of excitations (fermions) and can be block-diagonalised in each fixed particle-number sector. Moreover, because of the quadratic nature of the Hamiltonian, only the spectrum in the single-excitation subspace is needed in order to retrieve the full energy spectrum. We report here, for the sake of completeness, the mains steps for the derivation of the many-body dynamics in terms of single-body dynamics for non-interacting fermions, which is standard procedure in the second-quantization formalism.

The diagonalized form of the Hamiltonian  in Eq.~\ref{E.Ham_fer} in the single-excitation sector reads
\begin{equation}
\label{E_Ham_dia}
\hat{H}=\sum_{k=1}^N E_k\opkb{E_k}{E_k}
\end{equation}
where $\left\{E_k,\ket{E_k}\right\}$ are the eigenvalues and eigenvectors of the $N$-dimensional adjacency matrix of the graph with entries $J_{ij}$. Expressed in the position basis, $\ket{n}\equiv\hat{c}_n^{\dagger}\ket{\mathbf{0}}=\ket{00\dots1_n00\dots}$, where $\ket{\mathbf{0}}$ represents the fermionic vacuum state and $\ket{1_n}$ denotes the presence of a fermion on site $n$, the energy eigenstates in the single-excitation sector read $\ket{E_k}=\sum_{n=1}^{N}a_{kn}\ket{n}$, with $a_{kn}=\langle{n}\ket{E_k}$. The single-particle transition amplitude of an excitation from site $s$ to site $r$ is given by
\begin{equation}
\label{E.f1}
f_s^r(t)=\bra{r}e^{-i\hat{H}t}\ket{s}=\sum_{k=1}^N a_{r,k}a^*_{s,k}e^{-i E_kt}~.
\end{equation}
Because of the non-interacting nature of the Hamiltonian in Eq.~\ref{E.Ham_fer}, the energy eigenstates in the Hilbert space with $m$ fermionic excitations are given by
\begin{equation}
\label{E_neig}
\ket{E_{k_1k_2\dots k_m}}=\sum_{n_1<n_2<\dots<n_m=1}^{N}a_{n_1n_2\dots n_m,k_1k_2\dots k_m}\ket{n_1n_2\dots n_m}~,
\end{equation} 
with eigenvalues $E_{k_1k_2\dots k_m}=E_{k_1}+E_{k_2}+\dots+E_{k_m}$ and $a_{n_1n_2\dots n_m,k_1k_2\dots k_m}$ denoting the Slater determinant.

The many-body transition amplitude of $m$ excitations from sites $\mathbf{s}=\{s_1,s_2,\dots,s_m\}$ to sites $\mathbf{r}=\{r_1,r_2,\dots,r_m\}$ is readily obtained as a determinant of a matrix whose entries are the single-particle transition amplitudes in Eq.~\ref{E.f1}, 
\begin{align}
\label{E.fN}
&f_{\mathbf{s}}^{\mathbf{r}}(t)=\bra{\mathbf{r}}e^{-i\hat{H}t}\ket{\mathbf{s}}
=\sum_{k_1<k_2<\dots<k_m=1}^N e^{-i\left(E_{k_1}+E_{k_2}+\dots+E_{k_m}\right)t} \langle r_1r_2\dots r_m\ket{E_{k_1k_2\dots k_m}}\bra{E_{k_1k_2\dots k_m}}s_1s_2\dots s_m\rangle\nonumber\\
&=\begin{vmatrix}
f_{s_1}^{r_1}(t) & f_{s_1}^{r_2}(t) & \cdots & f_{s_1}^{r_m}(t)\\
f_{s_2}^{r_1}(t) & \cdots & \cdots & f_{s_2}^{r_m}(t)\\
\vdots & & \ddots &  \vdots\\
f_{s_m}^{r_1}(t) &  & \cdots & f_{s_m}^{r_m}(t)\\
\end{vmatrix}~.
\end{align}
The expression given in Eq.~\ref{E.fN} holds for every fermionic quadratic model, whereas if the operators in Eq.~\ref{E.Ham_fer} represent bosons, then, instead of the determinant, the many-body transition amplitude is given by the permanent of the matrix.

\section{The model}\label{Sec.1}

In this section, we apply the formalism of Sec.~\ref{Sec.MBD} to determine the two-excitation transition probability from a sender block to a receiver block, both composed of two sites, that are connected to a linear chain. The aim is to derive the conditions for the routing of the two excitations from the senders' to the receivers' location.  We will analyse two networks: a.) the receiver blocks have switchable couplings to the wire (Fig.~\ref{F.model}); b.) the receiver blocks are permanently coupled to the wire and the two sites in the sender block interact via a tuneable coupling (Fig.~\ref{F.model1}).

We consider Hamiltonians of the type given in Eq.~\ref{E.Ham_fer}, which, decomposed into the different components of the network, i.e., sender $S$, wire $W$, and receivers $R$, read
\begin{equation}
\label{E.Ham_gen}
\hat{H}=\hat{H}_S+\sum_i\hat{H}_{R_i}+\hat{H}_W+\hat{H}_{SW}+\sum_i\hat{H}_{R_iW}~.
\end{equation}

The Hamiltonian of the sender block and the $i$-th receiver block are, respectively
\begin{equation}
\label{E_HamS}
\hat{H}_S=J_s\left(\hat{c}_1^{\dagger}\hat{c}_{2}+h.c.\right)~,~
\hat{H}_{R_i}=J_i\left(\hat{c}_{r_i}^{\dagger}\hat{c}_{r_i+1}+h.c.\right)~,
\end{equation}
with $r_i$ denoting the position on the graph which will be given in the following.
The Hamiltonian for the quantum data bus reads
\begin{equation}
\label{E_Ham}
\hat{H}_w=J\sum_{n=1}^{n_w-1}\left(\hat{c}_n^{\dagger}\hat{c}_{n+1}+h.c.\right)~.
\end{equation}
Finally, the coupling between the sender block and the data bus site is assumed to be in the weak-coupling regime, $J_0\ll J,J_s, J_i$
\begin{equation}
\label{E_HamSC}
\hat{H}_{Sw}=J_0\left(\hat{c}_2^{\dagger}\hat{c}_3+h.c.\right)~;
\end{equation}
as well as the coupling between the $i$-th receiver block at location $r_i$ and the corresponding data bus site $w_i$, where $1\leq w_i\leq n_w$
\begin{equation}
\label{E_HamRC}
\hat{H}_{R_iw}=J_0\left(\hat{c}_{r_i}^{\dagger}\hat{c}_{w_i}+h.c.\right)~.
\end{equation}

For case a.) all couplings between the receiver blocks and the wire are switched off but one, embodying the recipient of the routing protocol and we set $r_i=n_w+2$, see Fig.~\ref{F.model} for an instance of the numbering choice of the sites following the sender-wire-receiver order. For  case b.) we follow the same ordering and an instance is given in Fig.~\ref{F.model1}.

The whole system sender+wire+receivers is made up of $N=n_w+2(n_r+1)$ sites with $r$ denoting the number of receiver blocks.


From Eq.~\ref{E.fN}, the two-body transition probability, with $\mathbf{s}=\{s_1,s_2\}$ and $\mathbf{r_i}=\{r_{i,1},r_{i,2}\}$, is given by 

\begin{align}
\label{E.3ampl}
|f_{\mathbf{s}}^{\mathbf{r_i}}(t)|^{2}=\left|\bra{1,2}e^{-it \hat{H}}\ket{N-1,N}\right|^2 = 
\begin{vmatrix}
f_{1}^{N-1}(t) &  f_{1}^{N}(t)\\
f_{2}^{N-1}(t) & f_{2}^{N}(t)\\
\end{vmatrix}^2~.
\end{align}

For only one sender and one receiver block located at opposite edges of the quantum wire, the model is one-dimensional and, using the Jordan-Wigner mapping from spinless fermions to spin-$\frac{1}{2}$ particles, the Hamiltonian in Eq.~\ref{E.Ham_gen} with open boundary conditions is equivalent to the $XX$ spin-$\frac{1}{2}$ model with nearest-neighbor coupling.

\begin{equation}\label{E.Spin_1D}
\hat{H}=\sum_{n=1}^{N}\frac{J_n}{2}\left(\hat{\sigma}^x_n\hat{\sigma}^x_{n+1}+\hat{\sigma}^y_n\hat{\sigma}^y_{n+1}\right)~.
\end{equation}

In such a case, it has been shown that two-qubit quantum state transfer~\cite{doi:10.1142/S021974991750037X, 1402-4896-2015-T165-014036} as well as entanglement generation of two Bell states~\cite{Apollaro2019a} is achieved with high fidelity. Modifications of the one-dimensional geometry have been investigated too. In Refs.~\cite{Vieira2019, VIEIRA20182586} each spin of the sender (receiver) block is coupled to the edges of the 1D quantum wire allowing for the transfer of a Bell state when operating in the single-excitation subspace. A similar geometry is adopted in Refs.~\cite{Yousefjani2020, Yousefjani2020a} with multiple sender (receiver) non-interacting spins coupled to the wire at the edges.

\subsection{Routing with switchable weak couplings}\label{S.switch}

\begin{figure}[H]
	\centering
	\includegraphics[width=\textwidth]{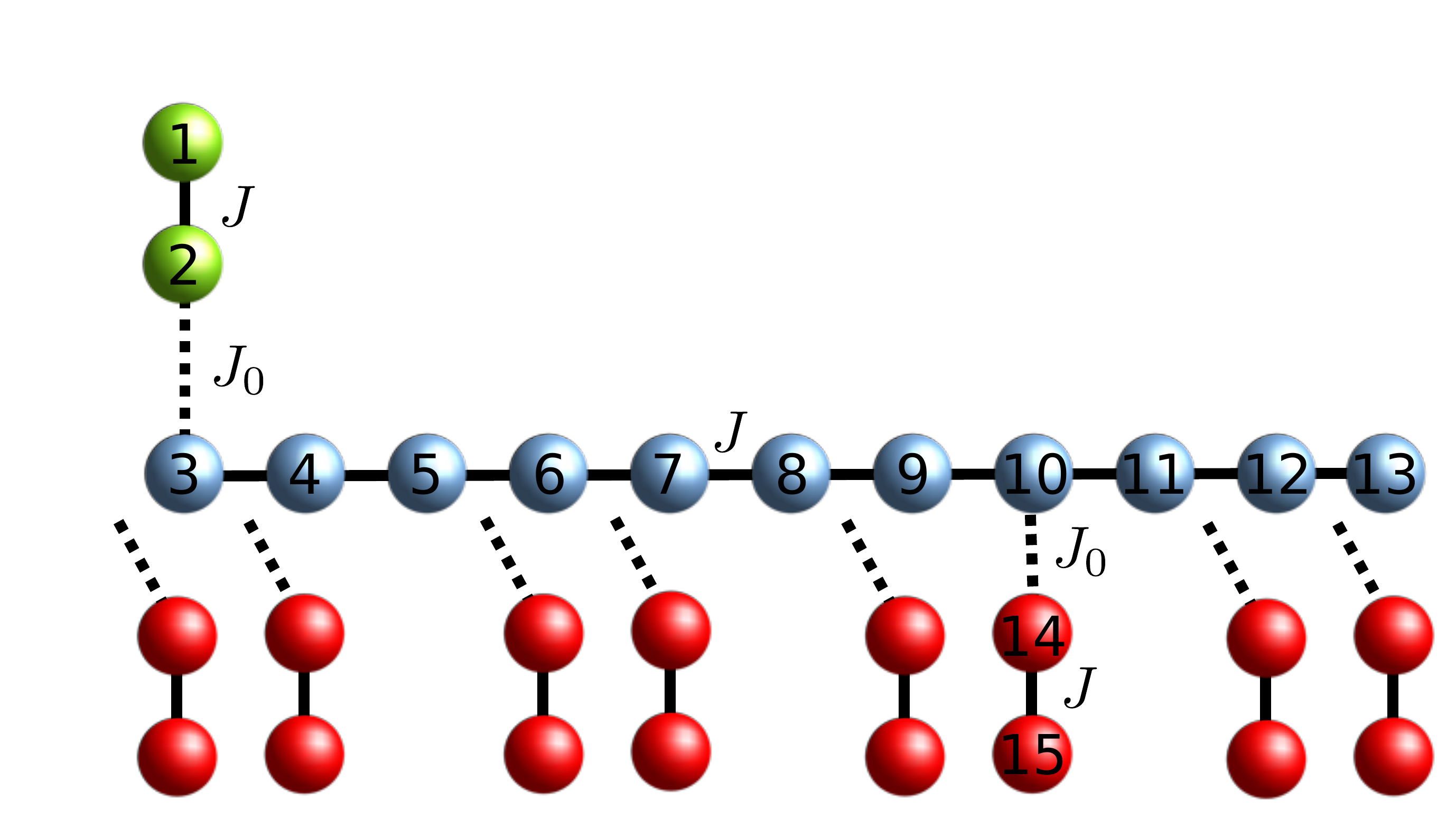}
	\vspace{0.2cm}
	\caption{Quantum routing of excitations by means of a linear chain quantum data bus with switchable interactions. The sender and receiver sites are depicted in green and red, respectively, while the quantum data bus sites are in blue. Continuous lines represent permanent couplings $J=1$, while dotted lines encode switchable weak couplings $J_0\ll 1$. Also shown is the numbering choice of the sites position adopted in Section~\ref{S.switch}.}
	\label{F.model}
\end{figure} 

Here we consider only one receiver block coupled to the wire for each execution of the routing protocol and, as we shall see, this allows us to assume uniform coupling within each component of the setup, i.e, the sender, the wire and the receiver blocks. We choose $J_s=J=J_r=1$ as our energy and time unit. On the other hand, the couplings between the sender (receiver) block and the wire will be in the weak-coupling regime, which we set throughout the paper to $J_0=0.01$.

The 1D-case where only one block of senders and one of receivers is each coupled at the edge of the quantum wire has been addressed in Ref.~\cite{Chetcuti2020}. There it has been shown that, although each length of the quantum wire $n_w$ allows for high-fidelity excitation transfer, for $n_w=3l+2$ ($l=0,1,2,\dots$), resonances between the sender (receiver) and the wire single-particle energy levels give rise to a faster transfer with respect to the instances $n_w=3l,3l+1$ where off-resonant transfer takes place.
In the former case, the single-particle transfer occurs on a time scale of the order of $J_0^{-1}$, yielding to a two-excitation transfer time scale of the order of $10~J_0^{-1}$ with the reason for the multiplicative factor being that the transfer dynamics involves a difference between eigenenergies that are perturbed to first-order in $J_0$. On the other hand, for the off-resonant dynamics, the two-particle transfer time is of order $J_0^{-2}$.  Considering that the excitation transfer mechanism holds in the perturbative regime $J_0\ll 1$, this may translate in severals of magnitude.

\begin{figure}[H]
	\centering
	\includegraphics[width=0.45\textwidth]{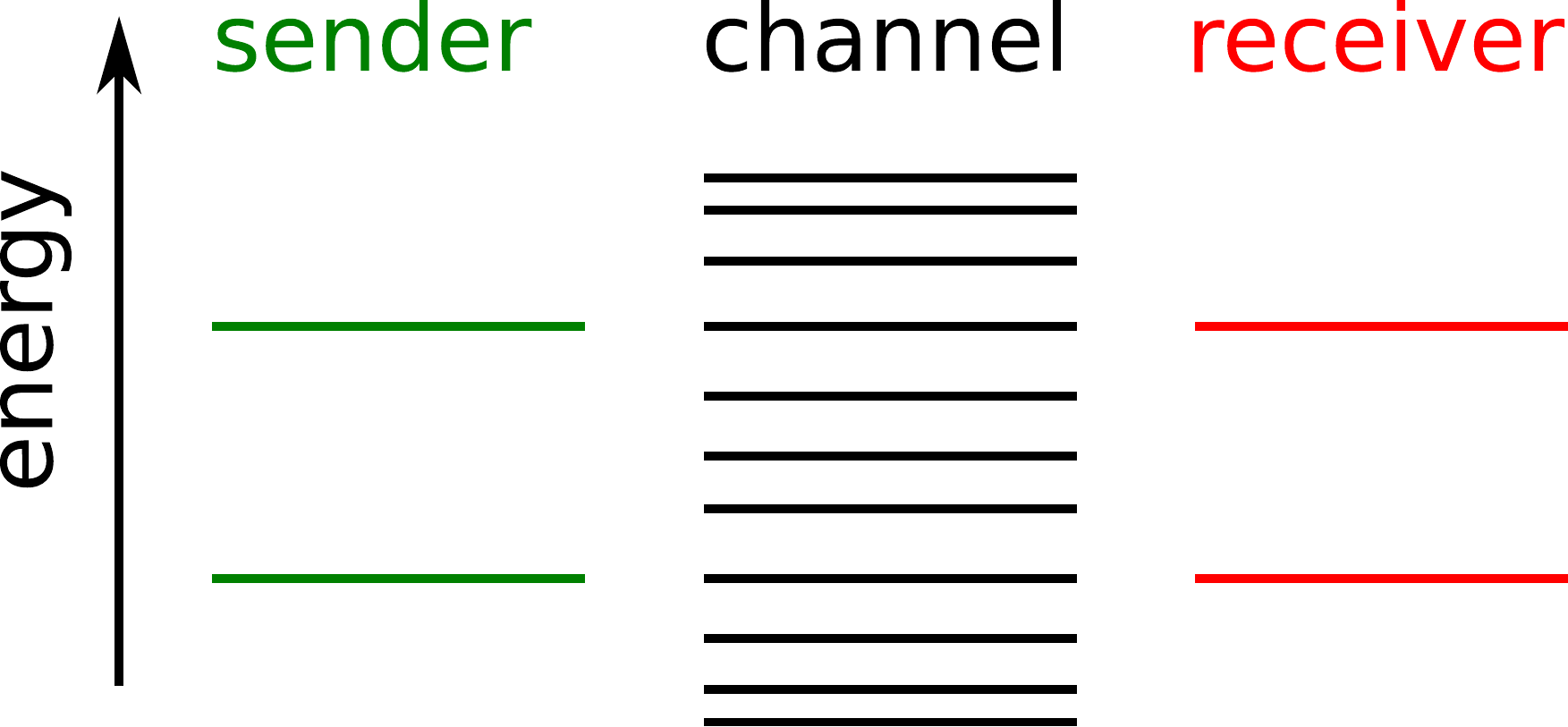}
	\hspace{5mm}
	\includegraphics[width=0.45\textwidth]{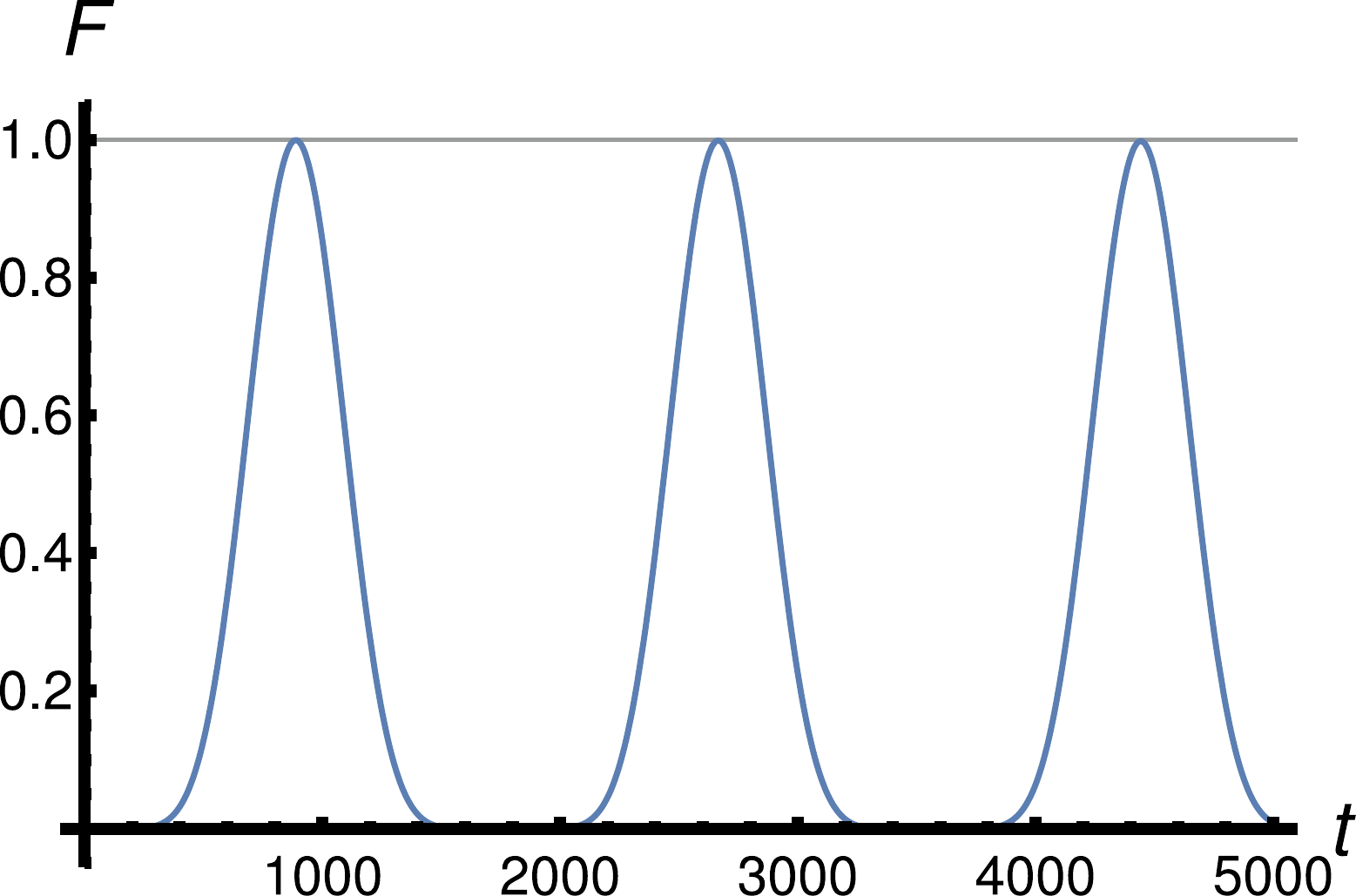}
	\vspace{0.2cm}
	\caption{(left) Single-excitation energy levels in the switchable coupling configuration. (right) Two-excitation transfer fidelity in the switchable configuration of Sec.~\ref{S.switch} with $n_w=11$, $r=7$, and $J_0=0.01$.}
	\label{F.swi}
\end{figure}

Here we address the case where the receiver block is coupled to the quantum wire at a different position $w_i$ than the edge of the chain, see Figure~\ref{F.model}. We also omit the suffix $i$ since only one receiver block is present in this protocol. We aim at finding the conditions on the position $w$ for which resonant transfer of the excitations from the sender to the receiver block at takes place. Following the argument for faster (resonant) transfer in Ref.~\cite{Chetcuti2020}, we set the length of the quantum wire $n_w=3l+2$ so that perturbative transfer is achieved for $w=n_w$, i.e., a receiver block can be coupled to the edge of the wire. For this wire length, we find that it is possible to couple a receiver block at each site $w\neq 3p$ of $n_w$, with $p$ integer. The fact that these latter sites of the wire cannot act as connection points for the receiver block can be explained by looking at the eigenstates of the wire's Hamiltonian $\hat{H}_W$ (Eq.~\ref{E_Ham}) that are resonant with the eigenstates of the sender (receiver) block. For $J=1$, the unperturbed energy level of the sender (receiver) is $E_{res}=\pm 1$, and, because of the mirror-symmetry of the Hamiltonians in Eq.~\ref{E_HamS}, they have the identical (absolute value) overlap on each site~\cite{Banchi2013} so that it suffices to consider only one of them.  The unperturbed energy levels of the wire that are resonant with the sender (receiver) are given by $E_k=2\cos \frac{k\pi}{n_w+1}=1$. Therefore, we obtain that, ordering $E_k's$ in decreasing order, the $k=\frac{n_w+1}{3}$-energy level of the wire is the resonant one, see left panel of Fig.~\ref{F.swi} for a schematic representation of the resonance condition. Expressing the corresponding energy eigenstate in the position basis
\begin{equation}
\label{E_no_point}
\ket{E_{res}}=\sqrt{\frac{2}{n_w+1}}\sum_{m=1}^{n_w}\sin \frac{k m \pi}{n_w+1}\ket{m}=\sqrt{\frac{2}{n_w+1}}\sum_{m=1}^{n_w}\sin \frac{m\pi}{3}\ket{m}~,
\end{equation}
meaning that the resonant energy level has no support on any site of the wire being multiple of 3. As a consequence, at first-order perturbation theory, the resonant energy level does not overlap with the receiver sites coupled to each third site of the wire, making the latter not apt as connection points for a two-excitation transfer. Furthermore, the $k_{res}$ eigenenergy state has constant spatial overlap with every other site $m\neq 3p$. This translates into a symmetric spatial distribution of the first-order perturbed eigenstates on the sender and receiver block, thus enabling the excitation transfer. Hence, for a wire of length $n_w=3l+2$ with uniform couplings equal to those within the sender (receiver) block, a total of $n_r=2\left(l+1\right)$ receiver points are possible. In Figure~\ref{F.swi} an instance of such a protocol is shown for $l=3$ with the receiver pair $n_{r_6}$ coupled to the quantum wire. In the right panel of Fig.~\ref{F.swi} an instance of the Rabi-like oscillations are shown for a wire's length of $n_w=11$ and connection point of the receiver block at $w=7$. We found in our numerical simulations for lengths of the wire in the order of the hundreds, that also for longer chains the fidelity reaches $F=1-O(J_0)$ for a receiver block connected at $w\neq 3p$ with the first peak of the oscillations occurring at a time of order $10~J_0^{-1}$.


%
%

\subsection{Routing with permanent weak couplings}\label{S.perm}
\begin{figure}[H]
	\centering
	\includegraphics[width=\textwidth]{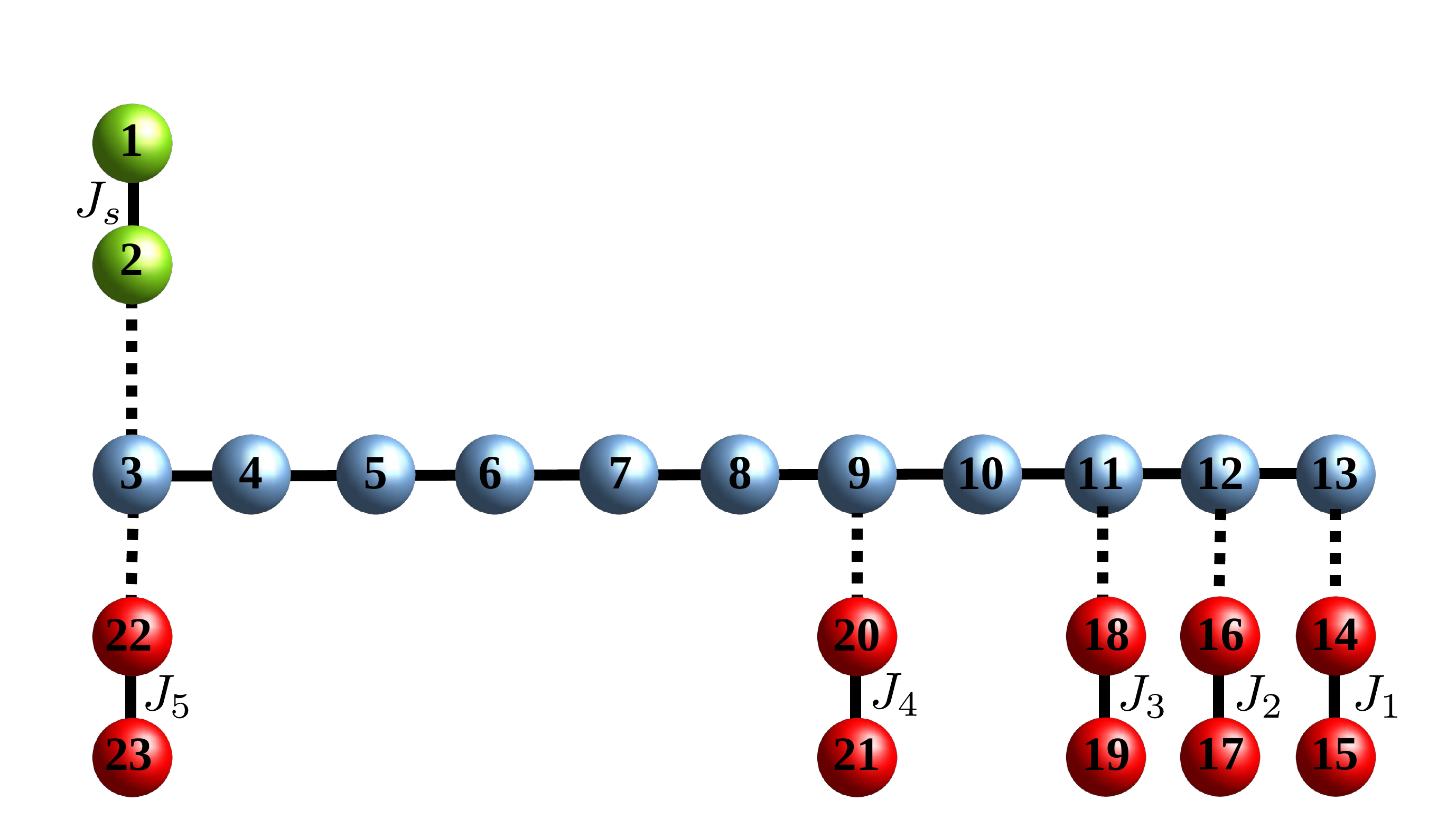}
	\vspace{0.2cm}
	\caption{Quantum routing of excitations by means of a linear chain quantum data bus with permanent interactions and assuming $J_s$ tuneable. The sender and receiver sites are depicted in green and red, respectively, while the quantum data bus sites are in blue. Also shown is the numbering choice of the sites position adopted in Section~\ref{S.perm}.
	}
	\label{F.model1}
\end{figure} 

A much more desirable routing configuration would be one without the need of switching on and off the couplings as described in the previous section. For the routing of a single particle, this has been achieved in Ref.~\cite{Paganelli2013} where in both the linear and the circular geometry, the sender chooses the receiver site tuning the only single-energy level on resonance with the receiver (and the quantum wire) energy level by means of a local magnetic field, i.e., the value of the local magnetic field acting on the receiver qubit is the \textit{routing address}. In the case of a sender block embodied by two particles, the same strategy does not work as a local magnetic field produces an uniform shift of both of the two energy levels and the simple sinusoidal excitation dynamics is lost. However, it is still possible to perform resonant routing with a sender block of two sites by using the intraspin coupling, which results in a symmetric energy shrinking or dilatation of the two single-energy levels. In such a case, the \textit{routing address} of each receiver block is given by their intraspin coupling $J_r$, see Figure~\ref{F.model1}.

In Section~\ref{S.switch}, we have shown that, for $J_s=1$, the wire's $k=\frac{n_w+1}{3}$-energy level is resonant with the sender block, and allows for the transfer of the two excitations to a receiver block with intraspin coupling $J_r=J_s$ provided that the connection point along the wire is $w\neq 3p$. The very same argument can be applied by tuning $J_s$ to a different value so that the single-energy levels of the sender block $E_s=\pm 2 J_s$ are resonant with two (symmetric) energy levels of the wire.  In order to match the resonance condition, an integer solution for $k$ has to satisfy the following equation
\begin{equation}
\label{E.res_con}
J_s=\cos\frac{k \pi}{n_w+1}~.
\end{equation}
That is, the $k=\frac{n_w+1}{\pi}\arccos J_s$-th energy eigenvalue of the wire is resonant with the sender. E.g., for $J_s=\frac{\sqrt{3}}{2}$, $k=\frac{n_w+1}{6}$.
Hence, the allowed contact points $w_i$ along the wire have to fulfill the condition that the resonant eigenstate spatial component of the contact point of the sender has to be equal to that of the receiver's contact point, i.e.,
\begin{equation}
\label{E.cont_points}
\sqrt{\frac{2}{n_w+1}}\sin\frac{k_{res}s\pi }{n_w+1}=\sqrt{\frac{2}{n_w+1}}\sin\frac{k_{res}w_i\pi }{n_w+1}~.
\end{equation}

To conclude this section, we recap the main results for routing a pair of excitations across a wire with uniform couplings to a desired location, specifying the resonance conditions on the sender and receiver couplings, respectively $J_s$ and $J_{r_i}$ and the respective allowed contact points.

For a wire of length $n_w$, the possible communication parties are  $\frac{n_w}{2}$ for even length chain and $\frac{n_w-1}{2}$ for odd length ones. Setting the intrawire coupling $J=1$, the single energy levels for which first-order excitation transfer occurs are $E_k=2\cos \frac{k \pi}{n_w+1}$, $k=1,2,\dots, \frac{n_w}{2}$ ($k=1,2,\dots, \frac{n_w-1}{2}$ for odd length chains). Each $k$ determines the intraspin coupling of the sender block via the relation $J_{r_i}=2\cos \frac{k_i \pi}{n_w+1}$ and the possible contact points $w_i$ along the wire via 
\begin{equation}
\label{E.CPs}
\sin \frac{k_i s \pi}{n_w+1}=\sin \frac{k_i w_i \pi}{n_w+1}~.
\end{equation}
Assuming that the sender is attached to the first site of the wire $s=1$, and exploiting the periodicity of the $\sin$ function, $\left|\sin \alpha\right|=\left|\sin \left(\alpha\pm n\pi\right)\right|$, with $n$ integer, 
\begin{equation}
\label{E_cond}
\frac{k_i w_{i}}{n_w+1}=\frac{k_i \pi}{n_w+1}\pm n\pi\rightarrow
w_{i}=\left|1\pm\frac{n\left(n_w+1\right)}{k_i}\right|~.
\end{equation}
Finally, the allowed contact points $w_i$ are the integers $\in[1,n_w]$ satisfying Equation~\ref{E_cond}. An instance of these conditions is given in Figure~\ref{F.model1} for $n_w=11$ and the corresponding values of $J_{r_i}$ and $r_i$ are given in Table~\ref{Tab}.

\begin{table}[H]
	\caption{Values of the intraspin couplings for the receiver blocks and available wire's connection sites for the receiver block for a wire's length $n_w=11$.}
	\centering
	\begin{tabular}{ccc}
		\toprule
		$k$	& \textbf{$J_r$}	& \textbf{$w_i$}\\
		\midrule
		1	& $\frac{\sqrt{3}-1}{2}$		& 1,11\\
		2	& 1			& 1,2,4,5,7,8,10,11\\
		3 & $\sqrt{2}$& 1,3,5,7,9,11\\
		4 & $\sqrt{3}$& 1,5,7,11\\
		5&$\frac{\sqrt{3}+1}{2}$&1,11\\
		\bottomrule
	\end{tabular}
\label{Tab}
\end{table}


\begin{figure}[H]
	\centering
	\includegraphics[width=0.45\textwidth]{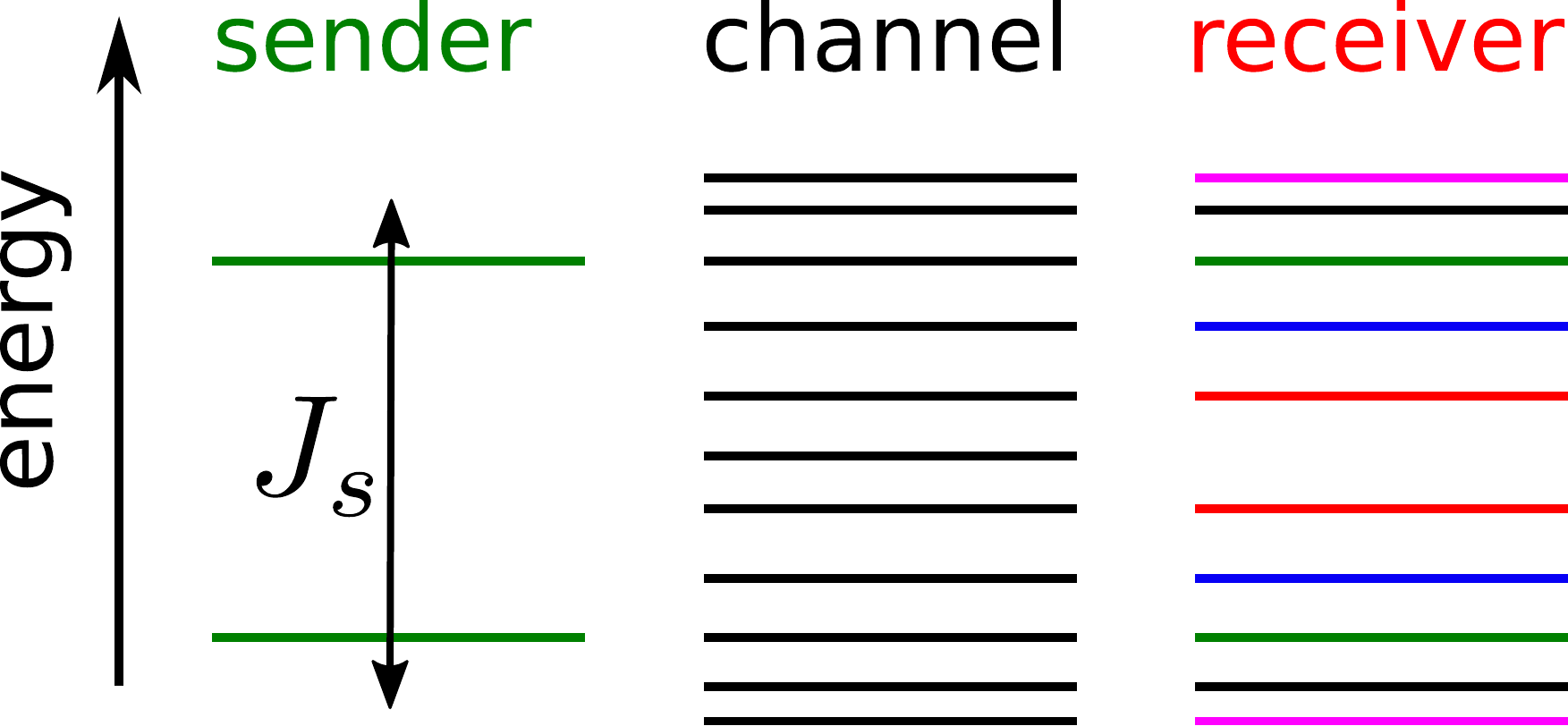}
	\hspace{0.5mm}
	\includegraphics[width=0.45\textwidth]{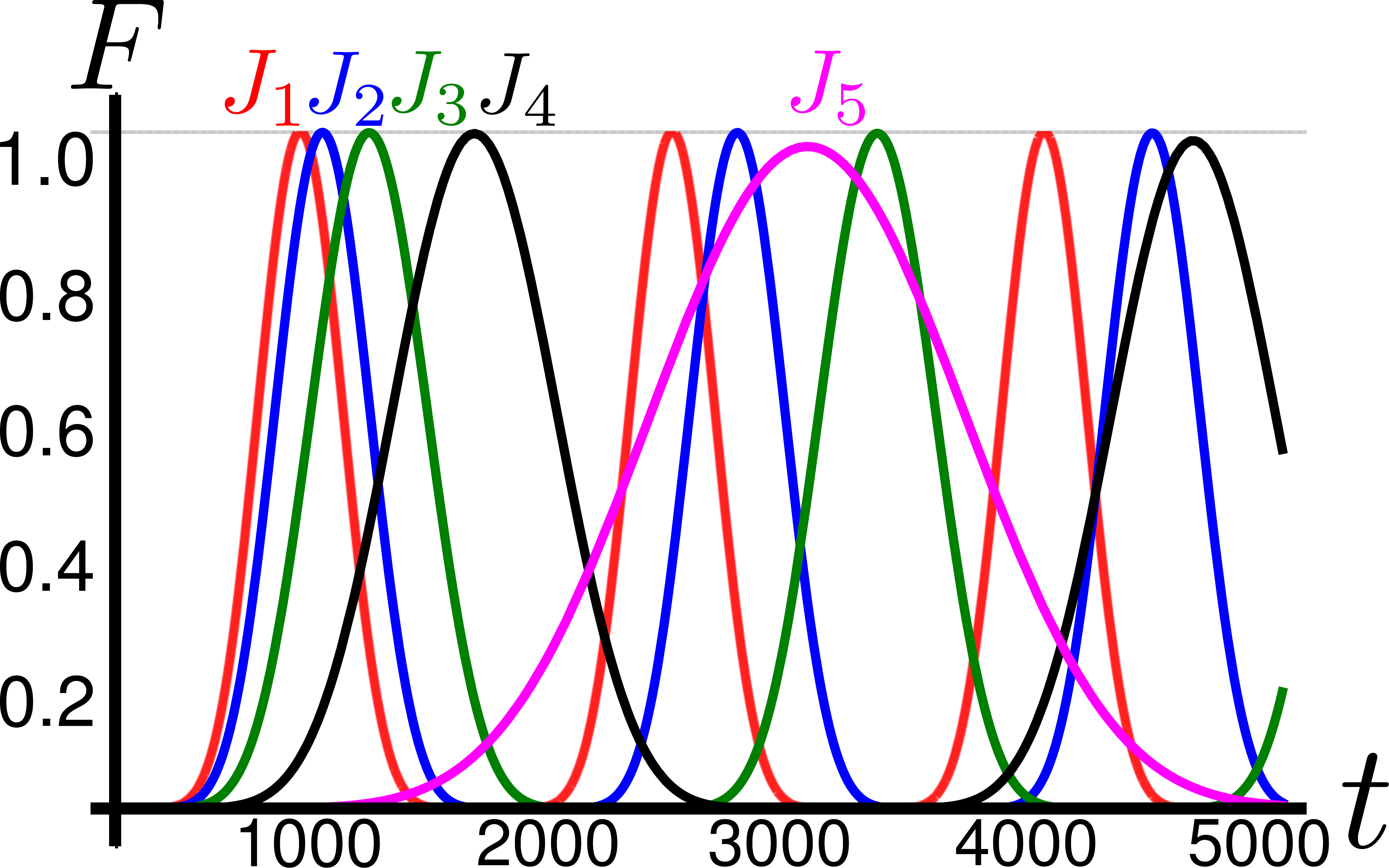}
	\vspace{0.2cm}
	\caption{(left) Single-excitation energy levels in the permanently coupled routing scheme. The sender's energy level can be tuned to be in resonance with a different pair of wire's (and receiver's) energy levels by tuning $J_s$. (right) Excitation transfer in the permanent coupling configuration of Sec.~\ref{S.perm} with $n_w=11$, $J_0=0.01$ and coupling scheme as in Fig.~\ref{F.model1}. The different curves correspond to the transfer fidelity of the two excitation to different receiver block by tuning $J_s$ to $J_{r_i}$. The colors of the curves correspond to the enegy levels in the left panel. }
	\label{F.linear_transfer}
\end{figure}
\section{Routing in spin systems}\label{S.Spin}
In the previous Sections, we have shown how, in the weak-coupling regime, routing of two-excitations from a sender to a receiver block can be achieved both in a switchable and a permanent coupling configuration in quadratic Hamiltonians. In this Section, we will consider the case when the network is made up of spin-$\frac{1}{2}$ particles interacting via an $XX$-type Heisenberg Hamiltonian. We will consider the switchable routing configuration depicted in Fig.~\ref{F.model} with Hamiltonian
\begin{equation}\label{E.Spin_Ham}
\hat{H}=\sum_{\langle ij \rangle}\frac{J_{ij}}{2}\left(\hat{\sigma}^x_i\hat{\sigma}^x_j+\hat{\sigma}^y_i\hat{\sigma}^y_j\right)~,
\end{equation}
where $\langle \rangle$ denotes the summation running over nearest-neighbor sites. Notice that Eq.~\ref{E.Spin_Ham} differs from Eq.~\ref{E.Spin_1D} because of the 2D nature of the network. By introducing the ladder operators $\hat{\sigma}^{\pm}=\frac{\hat{\sigma}^x\pm i \hat{\sigma}^y }{2}$, the Hamiltonian of the system can be obtained from Eqs.~\ref{E.Ham_gen}-\ref{E_HamRC} by substitution of $\hat{c}\rightarrow \hat{\sigma}^-$ and $\hat{c}^{\dagger}\rightarrow \hat{\sigma}^+$.
As already stated in Sec.~\ref{Sec.1}, were the network one-dimensional, i.e., the receiver block coupled to the last spin of the wire, then the Jordan-Wigner transformation would map Eq.~\ref{E.Spin_Ham} to a quadratic spinless fermion Hamiltonian. Such a case would constitute a special instance of the analysis in Sec.~\ref{S.switch} and several works on two-qubit quantum state transfer can be found in the literature. However, in the general case, where the receiver block is coupled to an arbitrary site of the wire, the Jordan-Wigner mapping does not apply as the system looses its one-dimensional nature. 

However, for a configuration such as the one depicted in Fig.~\ref{F.model}, there is only one spin belonging to the wire that has three nearest-neighboring spins; therefore, the one-dimensional nature of the model is only locally broken with the lowest possible coordination number. It is therefore interesting to investigate if the routing properties of the spinless non-interacting one-dimensional model of Sec.~\ref{Sec.1} still persist also when the network is made of spins when a rigorous mapping to fermions is not possible. 

Now, in order to evaluate the transition probability of two-excitations, we need to diagonalise the Hamiltonian in Eq.~\ref{E.Spin_Ham} in the two-excitations sectors, being the reduction to one-particle transition amplitudes not possible. The dimension of the Hamiltonian in the two-excitation Hilbert space is the binomial factor $\text{dim}\left[\hat{H}^{(2)}\right]=\binom{N}{2}$ and we diagonalise the Hamiltonian numerically for $N=306$ using the QuSpin package~\cite{Weinberg2019}.

\begin{figure}[H]
	\centering
	\includegraphics[width=\textwidth]{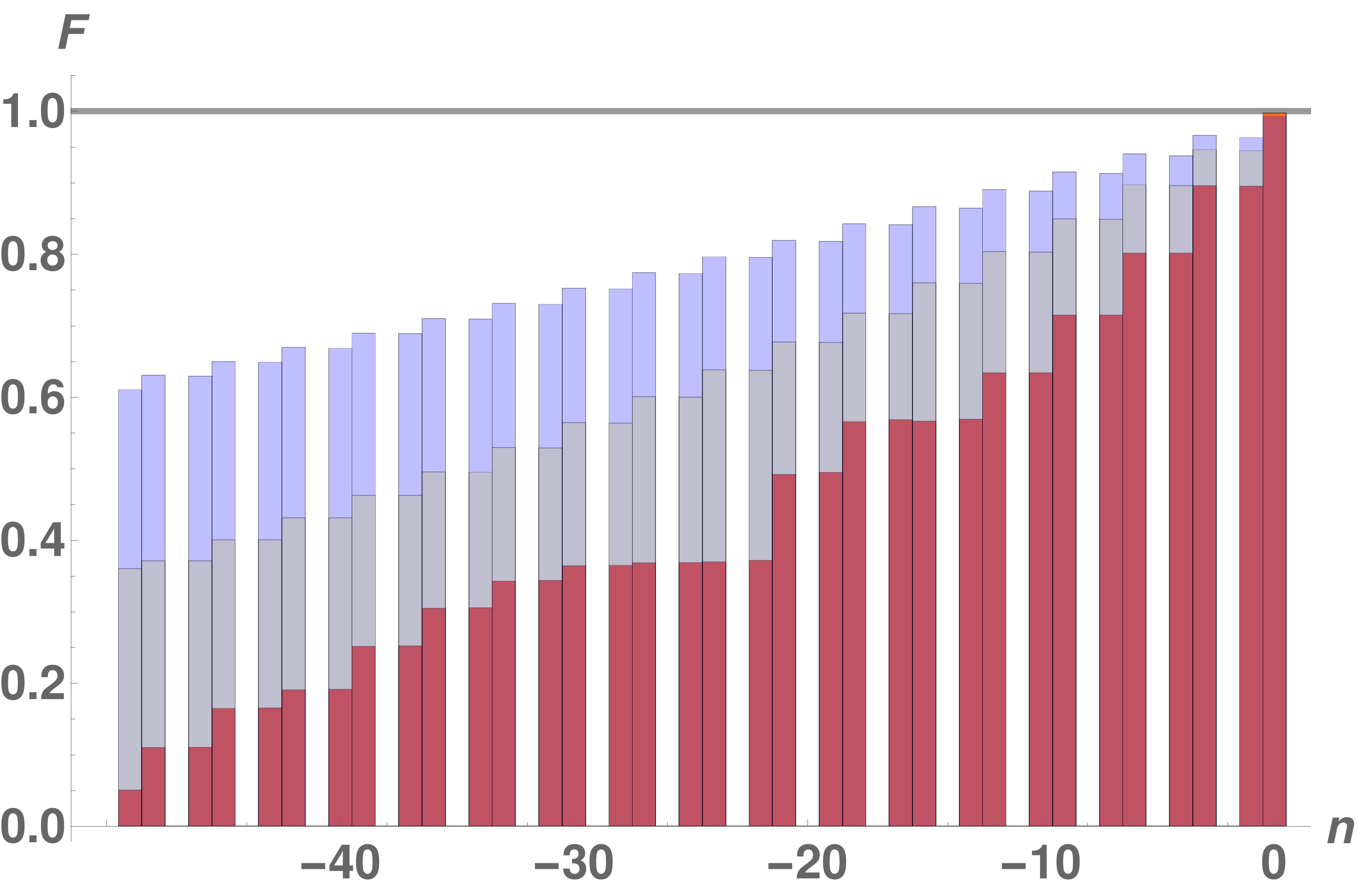}
	\caption{Transition probability for the switchable couplings protocol in Sec.~\ref{S.switch} with spin-$\frac{1}{2}$ particles sitting on the graph with $N=78$ (red), $N=156$ (gray), and $N=306$ (blue) and interacting via the $XX$ Hamiltonian in Eq.~\ref{E.Spin_Ham}. The numbering on the $x$-axis is the distance from the edge opposite the sender block $n=0,-1,-2,\dots$. Notice that, for the receiver block coupled to each third $n$ of the wire, the transiton probability is negligible. Interestingly, the quality of the transfer increases with the wire's length.}
	\label{F.spin}
\end{figure}

From Fig.~\ref{F.spin} we see that, as for the free-fermion network in Sec.~\ref{S.switch}, the transition probability of the two excitations from the sender block to the receiver block is negligible whenever the latter is coupled to every third spin of the linear chain. Furthermore, high-quality two-excitation transfer can be achieved, on a time-scale similar to that of the free-fermion network, only if the receiver block is coupled to connection points of the wire at the opposite edge with respect to the sender block. Moving away from that edge causes a linear decrease of the quality of the transfer with a lower slope the longer the wire. This may be seen as a consequence of the fact that the longer the wire, the more the one-dimensional nature of the system becomes manifest.

\section{Discussion}\label{S.Dis}
In this paper, we have investigated the routing of two fermionic excitations across a quantum network. In the proposed protocol we were able to show that two fermions, initially located on a sender block composed of two sites, can be efficiently routed to a receiver block of two sites, provided that both the former and the latter are weakly coupled to a one-dimensional quantum wire, modeled by a fermionic nearest-neighbor hopping Hamiltonian. We have proposed two different protocols: in the first one, we have assumed switchable couplings and derived the connection points of the wire which yields high-quality routing; in the second one we have assumed permanent couplings and envisaged in the tuneability of the sender's intrasite coupling a mean to route the two excitations to the desidered location. In each considered configuration we obtained a perturbatively-perfect fidelty, i.e., $F=1-O(J_0)$, where $J_0$ is the weak coupling of the sender and receiver block to the wire, with a transfer time scaling as $O(10 J_0^{-1})$. We also compared the fermionic network with a spin-$\frac{1}{2}$ network interacting via an $XX$-Heisenberg Hamiltonian. Due to the 2D nature of the network, the analysis had to rely on numerical evaluation and we found that, apart from the scenario where the receiver blocks are located towards the end of the wire, efficient routing is not achievable with qubits. However, our work hints towards the possibility to utilise very long quantum wires for the proposed 2-qubit routing protocol as we observed an enhancement of the routing fidelity by increasing the length of the wire.  

%
%

\vspace{6pt} 



\authorcontributions{
	All authors have contributed equally to the paper. All authors have read and agreed to the published version of the manuscript.}

\funding{This research has been carried out using computational facilities procured through the European Regional Development Fund, Project No. ERDF-080 “A supercomputing laboratory for the University of Malta}

\acknowledgments{We thank Kristian Grixti and Dr. Alessio Magro for technical support with the computational facilities.}

\conflictsofinterest{The authors declare no conflict of interest.}
%

\appendixtitles{no} 
%

\reftitle{References}


\bibliographystyle{unsrt.bst}





\end{document}